\documentclass[prl,aps,twocolumn,showpacs]{revtex4}
\usepackage{graphicx}
\usepackage{dcolumn}

\begin{document}

\title{Generic Phase Diagram of Fermion Superfluids with Population Imbalance}

\author{K. Machida}
\affiliation{Department of Physics, Okayama University,
Okayama 700-8530, Japan}
\author{T. Mizushima}
\affiliation{Department of Physics, Okayama University,
Okayama 700-8530, Japan}
\author{M. Ichioka}
\affiliation{Department of Physics, Okayama University,
Okayama 700-8530, Japan}
\date{\today}

\begin{abstract}
It is shown by microscopic calculations for trapped imbalanced Fermi superfluids that the gap function has always sign changes, i.e., the Fulde-Ferrell-Larkin-Ovchinnikov (FFLO) state like, up to a critical imbalance $P_c$, beyond which normal state becomes stable, at temperature $T=0$. A phase diagram is constructed in $T$ vs $P$, where the BCS state without sign change is stable only at $T\neq 0$. We reproduce the observed bimodality in the density profile to identify its origin and evaluate $P_c$ as functions of $T$ and the coupling strength. These dependencies match with the recent experiments.
\end{abstract}


\maketitle

Much attention has been focused on Fermion superfluidity realized experimentally using neutral Fermion atomic species, $^6$Li or $^{40}$K \cite{exp}. It is achieved by tuning the interaction strength via Feshbach resonance. 
Upon sweeping a magnetic field $H$ through the resonance situated at $H = 832 {\rm G}$ in $^6$Li case, 
the system exhibits a smooth crossover behavior from Bose-Einstein condensation at lower side to BCS at high field side. 

Recently two groups \cite{mit,rice} have succeeded in producing Fermionic superfluid in $^6$Li with the population imbalance where two interacting up and down spin species have different particle numbers. In attractive BCS side ($H\ge 832 {\rm G}$) which we focus on in this paper, the system shows a quantum phase transition as a function of the population difference, that is, the relative polarization $P \equiv |N_{\uparrow} - N_{\downarrow}|/(N_{\uparrow} + N_{\downarrow})$ 
Zwierlein {\it et al.} \cite{mit, private} assigned a critical imbalances $P_{\rm c} = 0.71$, beyond which normal state becomes stable, in the resonance  experiment at $H \simeq 832 {\rm G}$, and demonstrated: ($i$) $P_{\rm c}$ decreases as the temperature $T$ rises. ($ii$) $P_{\rm c}$ is proportional to the gap value $\Delta$, that is,  $P_{\rm c}\propto e^{-\pi/2k_F|a|} \propto \Delta$ ($k_F$ Fermi wave number). ($iii$) The spatial profile of the minority component exhibits a ``bimodal'' distribution, which disappears when the system becomes  normal either above $T>T_{\rm c}$ ($T_{\rm c}$  is the transition temperature) or $P>P_{\rm c}$. On the other hand, Partridge {\it et al.} \cite{rice} have found another transition at $P^{\ast}=0.09$ at the same resonance field (832G). 
 
The problem of the BCS with population imbalance posed by these experiments has been addressed in various contexts, ranging from ferromagnetic superconductor ErRh$_4$B$_4$ \cite{machida}, heavy Fermions superconductor CeCoIn$_5$ under a field \cite{kaku}, to color superconductivity in dense quark matter of high energy physics \cite{casalbuoni}. Among various proposals the Fulde-Ferrell-Larkin-Ovchinnikov (FFLO) state \cite{ff,lo} is a prime candidate to describe these experiments where the sign of the spatially varying  gap function $\Delta ({\bf r})$ changes periodically. This is contrasted with the usual BCS state in which $\Delta({\bf r})$ keeps a definite sign even though they might spatially vary in some situations. Here, we call it the  ``BCS'' state which has a definite sign in the gap function. The ``FFLO'' state changes its sign somewhere in the system, even though both states have spatially varying gaps due to trapping.

Prior to the present work, there have been papers devoted to this problem. However, most works are either an infinite uniform system without trapping \cite{uniform} or does not consider possibility of the FFLO state \cite{non-fflo,ho}, except for a few \cite{few}. It is crucial to take into account two effects simultaneously because we are considering an intrinsically non-uniform finite system. The energy difference of the FFLO and the BCS is so subtle because in the FFLO solution the sign change occurs only near the surface and both spatial profiles are similar at the trap center. 

The purpose of this paper are two-fold: (1) To construct a generic phase diagram in the plane of  $T$ versus $P$ and (2) to characterize each phase by the microscopic calculation considering a trap. In particular, the origin of the observed bimodality in minor component, item $(iii)$, can be attributed to a characteristic of the FFLO state. The derived phase diagram with the Lifshitz point \cite{chaikin} turns out to be quite universal where three second order lines meet at the tricritical Lifshitz point \cite{machida1984,fujita}.  
A new aspect here is that the population in two species is the control parameter while in usual condensed matter systems the relative shift of the chemical potential is controllable. The resulting phase diagram shows that at $T=0$ the FFLO state is ubiquitous and always the ground state for any imbalanced cases up to a critical value. It allows us to understand two critical values $P_c$ \cite{mit} and $P^{\ast}$ \cite{rice} at $T>0$ in addition to items $(i)$ and $(ii)$.

It is convenient to use the Bogoliubov-de Gennes (BdG) formalism to describe $\Delta ({\bf r})$ with population imbalance under a trap $V({\bf r})$. Throughout this paper, we set $\hbar=k_B = 1$. We consider a cylindrical system with $V({\bf r})={1\over 2}M\omega r^2$, imposing a periodic boundary condition with the periodicity $5d$ ($d^{-1} \equiv \sqrt{M\omega}$) along the $z$-direction. Thus we are treating a three dimensional (3D) system, depending on the radius $r$ and $z$. In the current work, the longitudinal trap along the $z$-axis is absent. The BdG equation for the quasi-particle wave functions $u_{\bf q}({\bf r})$ and $v_{\bf q}({\bf r})$ labeled by the quantum number ${\bf q}$  is read as follows with the local density of each spin state $\rho _{\sigma} (r)$ ($\sigma = \uparrow, \downarrow$) and attractive interaction $g$ \cite{mizushima}:
\begin{eqnarray}
\left[ 
	\begin{array}{cc}
		\mathcal{K}_{\uparrow} (r) & \Delta (r) \\
		\Delta ^{\ast} (r) & - \mathcal{K}^{\ast}_{\downarrow}(r) 
	\end{array}
\right] 
\left[ 
	\begin{array}{c} u_{\bf q}({\bf r}) \\ v_{\bf q}({\bf r}) \end{array}
\right] =  \varepsilon _{\bf q}
\left[ 
	\begin{array}{c} u_{\bf q}({\bf r}) \\ v_{\bf q}({\bf r}) \end{array}
\right],
\label{eq:bdg}
\end{eqnarray}
where $\mathcal{K}_{\uparrow,\downarrow} (r) = - \frac{\nabla^2_{\bf r}}{2M} +V(r) + g\rho _{\downarrow,\uparrow} (r) - \mu _{\uparrow,\downarrow}$ 
with the third term being the Hartree term.
The self-consistent gap equation is given by
\begin{eqnarray}
\Delta(r) = g\sum_{\bf q} u_{\bf q}({\bf r}) v^{\ast}_{\bf q}({\bf r}) f(\varepsilon _{\bf q}) - \frac{g}{2}\Delta(r)G^{\rm irr}_{E_F} (r). 
\label{eq:gap}
\end{eqnarray} 
Since the chemical potential shift $\delta \mu \equiv \mu _{\uparrow} - \mu _{\downarrow}$ causes the braking of the time-reversal symmetry, the sum in Eq.~(\ref{eq:gap}) is done for all the eigenstates with both positive and negative eigenenergies \cite{ichioka}. To regularize $\Delta(r)$, we have subtracted $G^{\rm irr}_{E_F}$ which is the irregular part of the single-particle Green's function \cite{bruun}. 

In Fig.~1 we show the phase diagram at $T=0$ in the plane of $P$ and the energy gap $\Delta_0 (0)$ at the trap center for equal population case normalized  by the Fermi energy $E_{F}(0)$ at the center. It is seen that the critical imbalance $P_{c}$ beyond which the normal state becomes stable is linearly proportional to $\Delta_0 (0)/E_{F}(0)$. This shows the critical chemical potential difference $\delta \mu _{c}$ is given by $\delta \mu _{c} /E_{F}= \alpha \Delta_0  /E_{F}$ where the proportional constant $\alpha$ delicately depends on the dimensionality of the system and/or the Fermi surface shape. For example,one can find $\alpha _{\rm 1D} \rightarrow \infty$ \cite{machida} in one dimension and $\alpha _{\rm 3D} \simeq 1.41$ \cite{takada} in 3D Fermi sphere. Since $P= \frac{3}{4}\frac{\delta \mu}{E_{F}}$ for the normal state with 3D Fermi sphere under the assumption $\delta \mu \ll E_F$, $P_c =\frac{3}{4}\alpha _{\rm 3D}\frac{\Delta_0}{E_{F}}\simeq 1.14 {\Delta_0 \over E_{F}}$, which is also shown in Fig.~1. This is changed in the presence of the harmonic trap to $P_c = \frac{3}{2}\alpha _{\rm ho}\frac{\Delta _0(0)}{E_{F}(0)}$ because $P=\frac{3}{2}\frac{\delta \mu}{E_F(0)}$ in the normal state. Our numerical calculation in Fig.~1 shows $\alpha _{\rm ho} \simeq 1.26$.

\begin{figure}[t]
\includegraphics[width=0.85\linewidth]{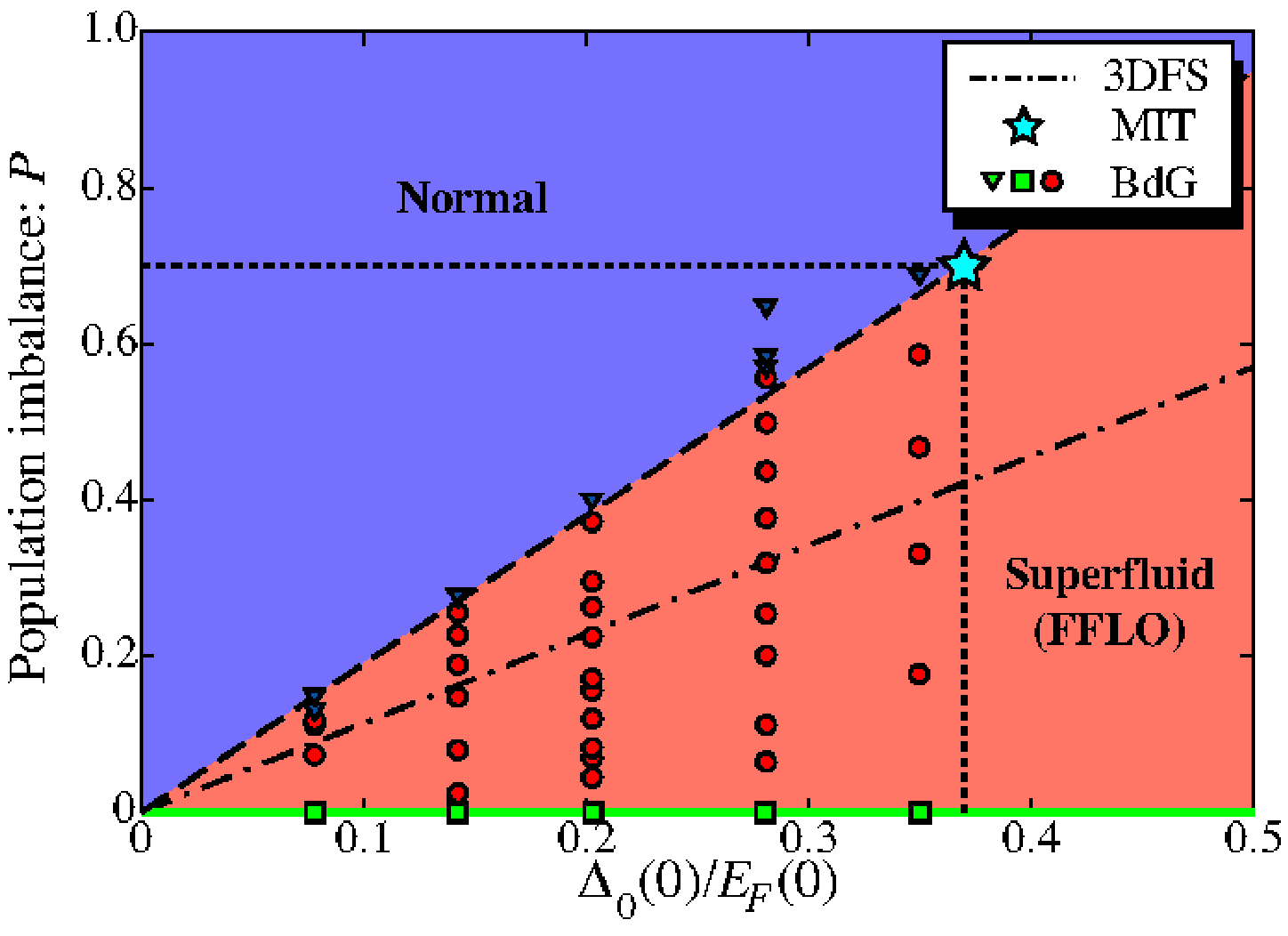} 
\caption{
(color online) Phase diagram in $P$ vs $\Delta_0(0)/E_F(0)$. Circles, squares, and triangles show the FFLO, BCS, and normal states obtained by our calculation, respectively. Dotted-dashed line denotes the result of 3D Fermi sphere case \cite{takada}. Star is the experimental date on the resonance ($H=832$G)  \cite{mit}.
}
\label{fig:gap}
\end{figure}

We note that the linear relation $P_c \propto {\Delta_0 (0)\over E_{F}(0)}$ is observed experimentally since the observed phase boundary between the normal and superfluid states is approximately exponential behavior, namely, $P_c\propto e^{-\pi/2k_F|a|}$ (see Fig.~5 in Ref.~\cite{mit}). This agreement must be checked further experimentally for wider $1/k_F|a|$ region.

This enhanced critical value $P_c$ given by $\delta\mu _c / \Delta _0 = \alpha _{\rm ho}$ far exceeds other known values, such as the so-called Pauli-Clogston limiting value $\delta \mu _c / \Delta_0 =1/\sqrt{2}$ signaling the first order transition from BCS to the normal state, or BCS-FFLO unstable point $\delta \mu _c/ \Delta_0 = 2/\pi$ which corresponds to the one-soliton creation energy\cite{machida}. This indirectly proves that the present solution of the BdG equation is stable energetically.

In order to obtain the observed $P_c = 0.7$ at the resonance \cite{mit}, we can read off from Fig.~1 that ${\Delta _0(0)\over E_F(0)} \simeq 0.38$, by performing the naive extrapolation from the weak coupling limit. This value is compared with the theoretical estimate ${\Delta_0 \over E_F} \sim 0.49$ by Bulgac {\it et al.} \cite{bulgac}. The direct estimation of $P_c$ at unitarity limit is still open to question. 

It is also shown in Fig.~1 that at $T=0$ in the stable superfluid state the order parameter $\Delta (r)$ always exhibits the sign change except for equal population ($P=0$), that is, the FFLO state is stable. The sign of $\Delta (r)$ must change to accommodate excess majority species at $T=0$, which is nothing but FFLO, while at $T \neq 0$ the ``magnetization'' can be accompanied with the non-oscillating pairing via thermally excited quasi-particles. The $\pi$-shift, i.e., sign change, of the gap function is essence of the ``topological doping'' important for stripes in high $T_c$ superconductors \cite{stripe}. This is also well-known in the other physics field \cite{mertsching}; The commensurate (C) charge or spin density waves (CDW, SDW) give way to the incommensurate (IC) ones when adding excess carriers. The present problem is precisely analogous to this C-IC problem, where IC (C) corresponds to FFLO (BCS). 


We are now in position to characterize the FFLO and BCS states. We show two typical examples of these states in Fig.~2. In FFLO shown in Fig.~2(a) we can divide the density distribution into three distinctive regions (I), (II), and (III) from the center. In ``BCS core'' region (I) the up and down-spin atoms are nearly equal and balanced. The population of down-spin atoms  is enhanced and pulled up by mutual attraction. The gap develops fully there and the magnetization $m({\bf r})=\rho_{\uparrow}({\bf r})-\rho_{\downarrow}({\bf r})$ almost vanishes. In ``FFLO'' region (II) the gap function changes its sign, allowing to accommodate the excess majority species. These features give rise to (A) the bimodal distribution in the minor component  (see a shoulder of $\rho_{\downarrow}(r)$ in lower panel of Fig.2(a)) and (B) a sharp peak structure in $m(r)$ at $r/d \sim 2.7$. These features which are observed experimentally \cite{rice,private} come from the sign changes of $\Delta(r)$. We note that the ordinary BCS region (I) is describable with the local density approximation \cite{non-fflo}, while the oscillating pairing in (II) results from only the full numerical calculation of the BdG equation (1). In ``complete polarization'' region (III) the gap is almost vanishing and there is no minority species. Thus the complete polarization is attained there.

\begin{figure}[t!]
\includegraphics[width=0.48\linewidth]{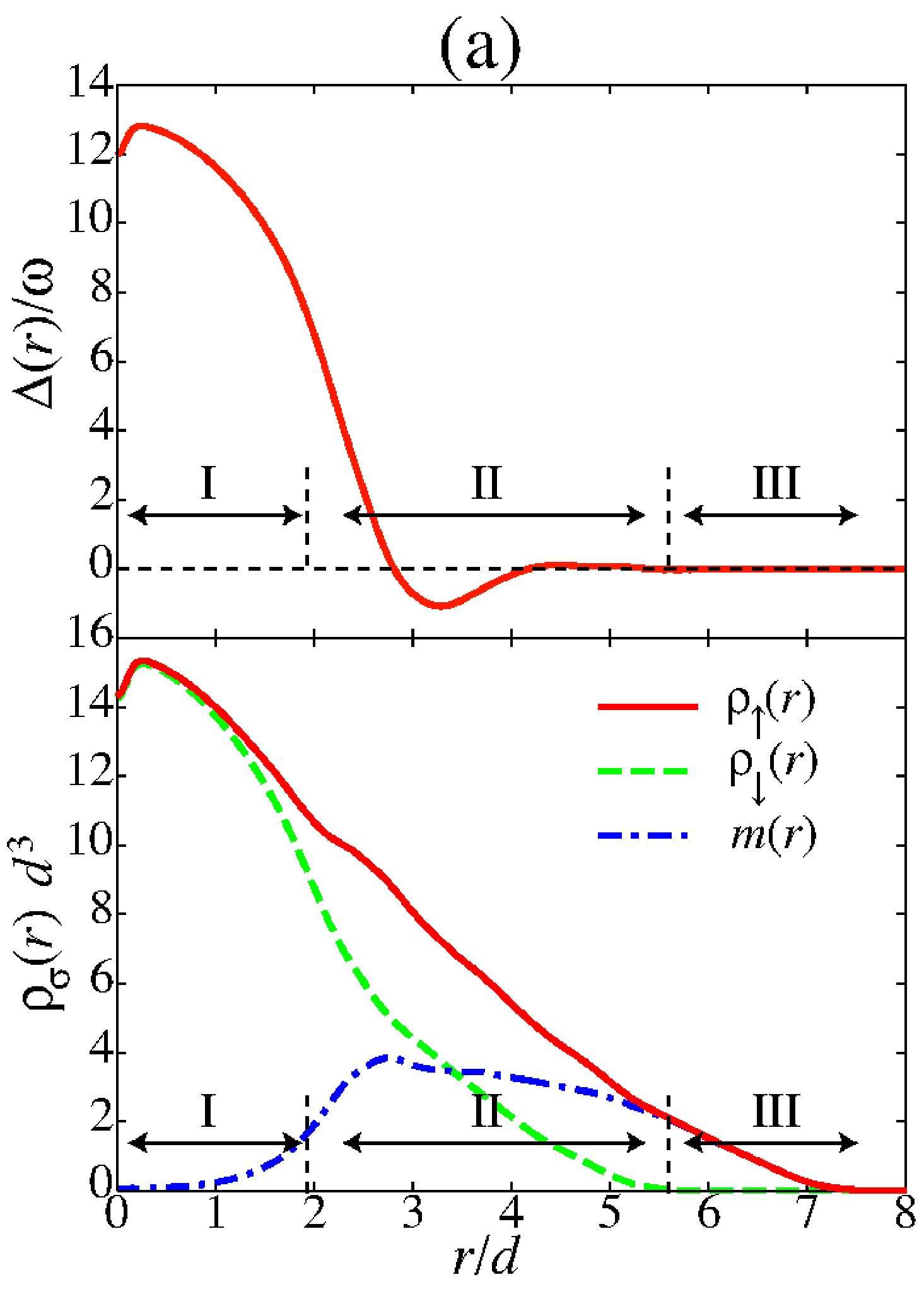} 
\includegraphics[width=0.48\linewidth]{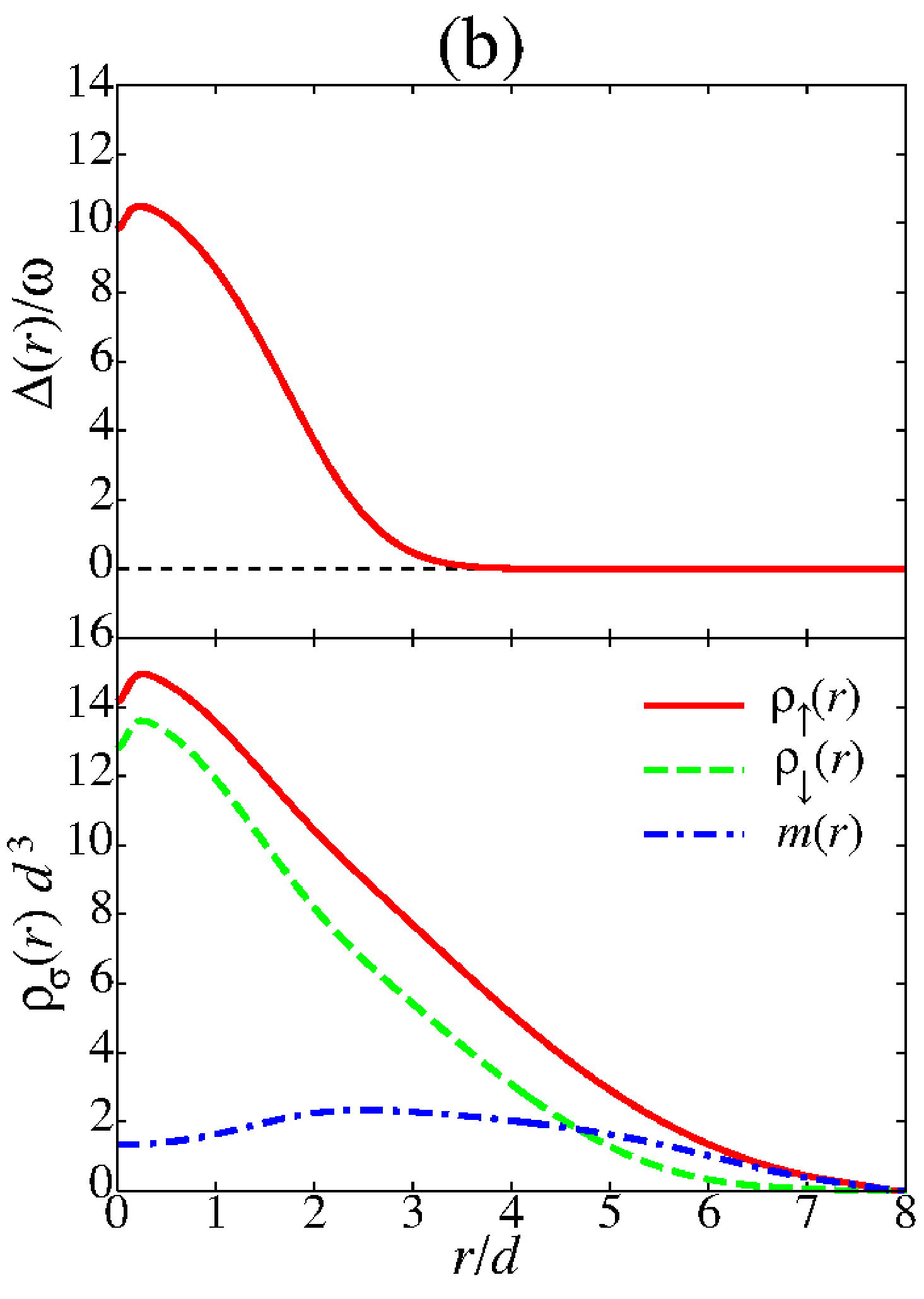} 
\caption{
(color online) Spatial distributions of $\Delta(r)$, $\rho_{\sigma}(r)$
and $m(r)$ for (a) FFLO state ($P=0.44$ and $T$=0)
and (b) BCS state ($P=0.22$ and $T/\omega$=4.0). $\Delta_0(0)/E_F(0)$=0.28. $\omega$ is the trap frequency.
}
\label{fig:gap}
\end{figure}

These characteristics (A) and (B) are indeed observed experimentally \cite{mit}. Note that experimental data \cite{mit}
 are obtained by the columnar integrated density distributions, yet they show prominent bimodality and sharp peak structures, implying that the actual three dimensional features are sharper. 

These characteristics in the FFLO are contrasted with those in the BCS shown in Fig.~2(b); The gap has a definite sign and the two density profiles for two species are smooth and scaled to each other. At $T>0$, the difference $\rho _{\uparrow}(r)-\rho_{\downarrow}(r)=m(r)$ appears by thermal excitations. The outer edge contains only the up-spin atoms where the gap vanishes. The resulting $m(r)$ has no sharp feature. The minority distribution ceases to  exhibit the bimodality. These features are almost the same as those in the normal state given by the Thomas-Fermi profile. These features are observed
either in the region $P>P_c=0.7$ in Ref.~\cite{mit} or $P<P^{\ast}=0.09$
in Ref.~\cite{rice}.

\begin{figure}[t!]
\includegraphics[width=0.75\linewidth]{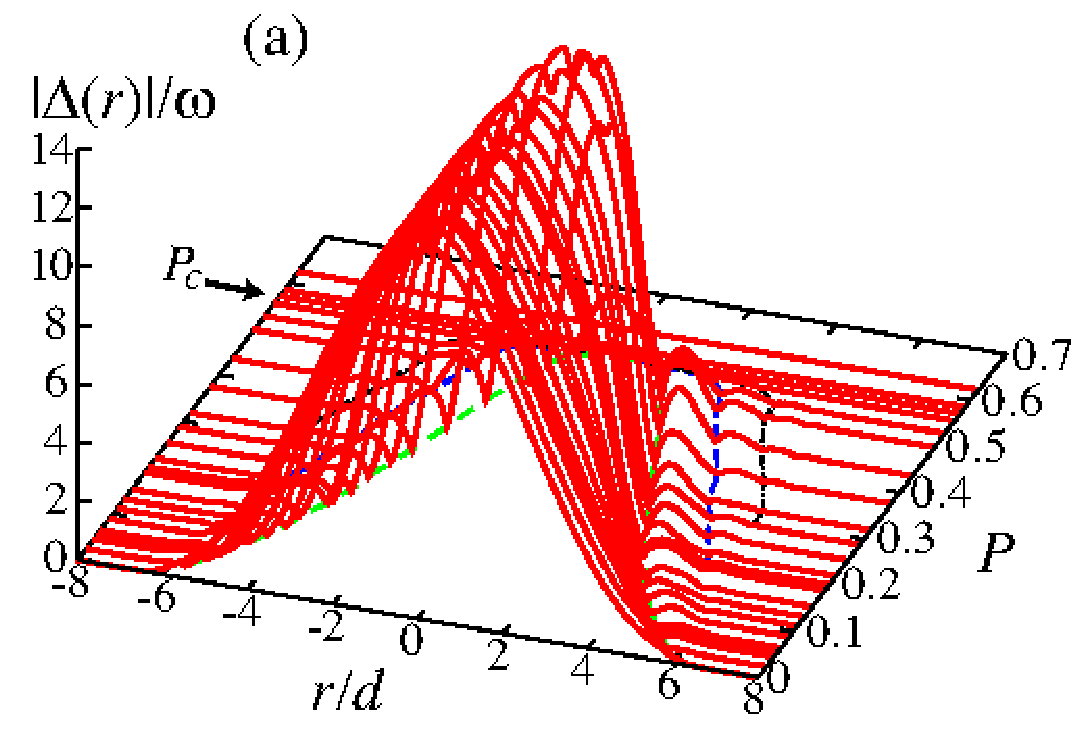} \\
\includegraphics[width=0.75\linewidth]{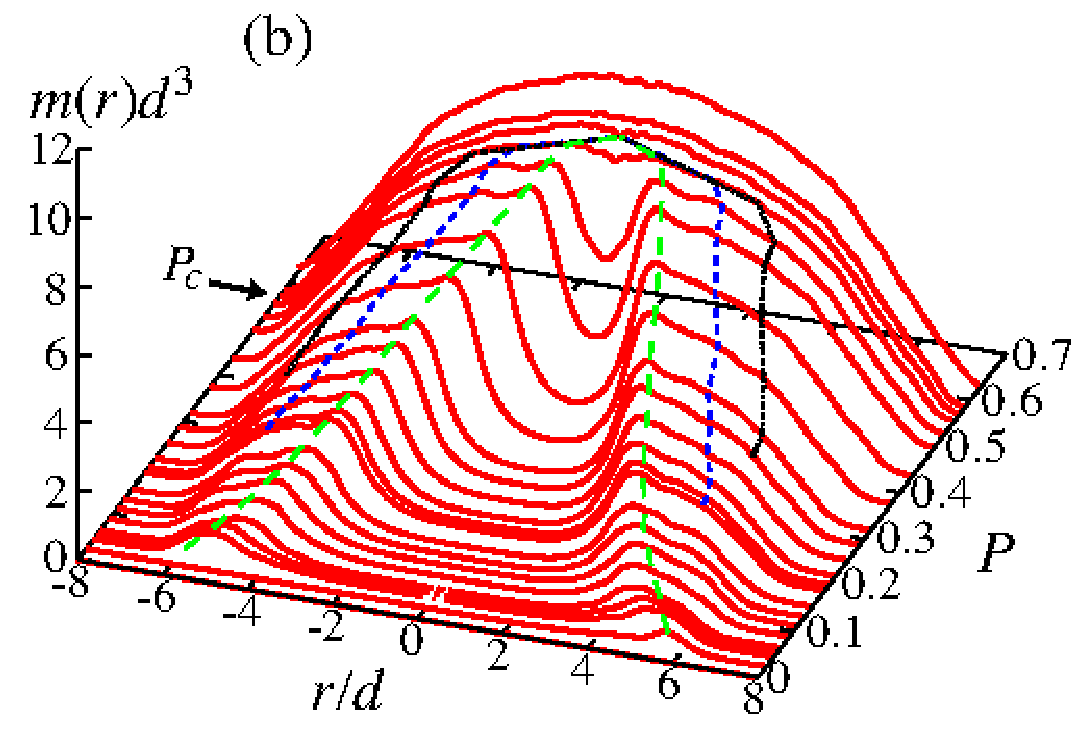} 
\caption{
(color online) Spatial distributions of $|\Delta(r)|$ (a) and $m(r)$ (b) for various $P$'s. $\Delta_0(0)/E_F(0)$=0.28 and $T=0$. The normal state appears for $P>P_c = 0.57$. The dotted-dashed lines show the nodes of $\Delta(r)$. 
}
\label{fig:gap}
\end{figure}

In order to better characterize the FFLO, we display series of changes both $|\Delta (r)|$ and $m(r)$ at $T=0$ in Fig.~3. It is seen that as $P$ increases, the periodicity of the oscillations in $\Delta (r)$ relatively stays constant and the number of the sign change increases. As for $m(r)$, with increasing $P$ the double peak structure changes into a single peak above $P_c$, signalling phase transition from the FFLO state to the normal state. We can see small and faint features in $m(r)$ corresponding to the sign change of $\Delta(r)$.

In Fig.4 we show the phase diagram in the plane of $T/T_{c0}$ versus $P/ P_{c0}$ where $T_{c0}$ is the transition temperature at $P= 0$ and $P_{c0}$ is the critical imbalance at $T=0$. This is determined by solving Eqs.(1) and (2) for various $T$ and $P$ where circle (square) shows the FFLO (BCS) state and inverted triangle is the normal state. All lines indicate the second phase transitions which meet at a tricritical point $L$ known as the Leung point, or  Lifshitz point in more general context \cite{chaikin}. The BCS-FFLO line starts right from $P=T=0$, implying that the ground state is always FFLO when $P\neq 0$ as mentioned above. The BCS appears only at a finite $T$. From Eqs.(1) and (2), it is easy to derive the equation for $T_c$, or the boundary between the superfluid and normal state as function of $P$, namely, $1 = g \int d{\bf r}\int d{\bf r}' \Psi^{\ast}(r)[K({\bf r},{\bf r}') -\delta({\bf r}-{\bf r}')G^{\rm irr}_{E_F}(r)]\Psi(r')$, where
\begin{eqnarray}
K({\bf r},{\bf r}') =
\sum _{{\bf q}, {\bf q'}}{f(\varepsilon _{\bf q})-f(\varepsilon _{\bf q'})\over{\varepsilon _{\bf q}-\varepsilon _{\bf q'}} }  u_{\bf q'}({\bf r})u^{\ast}_{\bf q'}({\bf r}')v^{\ast}_{\bf q}({\bf r})v_{\bf q}({\bf r}').
\label{eq:Tc}
\end{eqnarray}
The gap function $\Psi(r) \propto \Delta(r)$ is normalized as $\int d{\bf r} |\Psi(r)|^2= 1$. Using $\Psi(r)$ for the BCS state, we estimate $T_c$.  The results are also plotted in Fig. 4 as the dashed-dotted line, showing a good agreement  with the full numerical computation. We have confirmed that $T_c$ in FFLO becomes higher than that in BCS beyond the Lifshitz point, proving the stability of FFLO over BCS.

\begin{figure}[t!]
\includegraphics[width=0.8\linewidth]{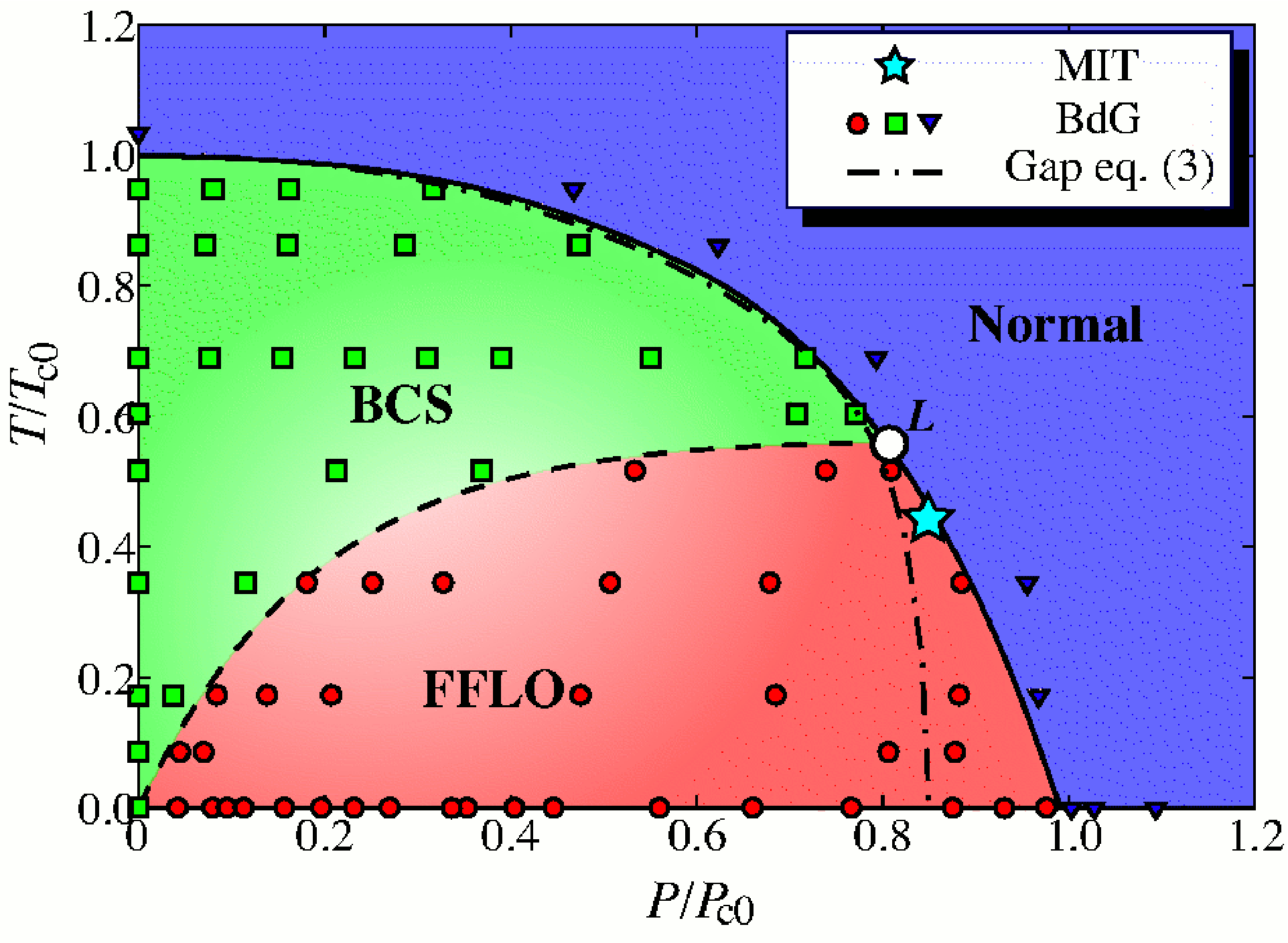} 
\caption{
(color online) Phase diagram in $T/T_{c0}$ vs $P/P_{c0}$ for FFLO (circle), BCS (square), normal state (triangle) at $\Delta _0(0)/E_F(0)=0.28$. The dashed-dotted line is $T_c$ for BCS obtained from Eq.~(\ref{eq:Tc}). Empty circle is the Lifshitz point. Star is the experimental data \cite{private}. 
}
\label{fig:gap}
\end{figure}

According to the experiment \cite{mit} $P_c(T)$ decreases as $T$ increases for three magnetic fields ($H=832$, 883, and 924G). More quantitatively, at the resonance, $P_c(T)/P_{c0}=0.86$ at $T/T_{F}=0.12$ \cite{private}, which is indicated in Fig.~4, estimated within the experimental value on the resonance \cite{kinast}: $T_{c0}/T_F=0.27$. These agree with our results. It is interesting to notice that under a fixed temperature, say, $T/T_{c0}=0.3$ as $P$ increases, BCS changes into FFLO at $P/P_{c0}= 0.1$ and upon further increasing $P$, FFLO finally becomes unstable at $P/P_{c0} \simeq 0.9$. It is reasonable that there are two transitions observed by Partridge {\it et al.} ($P^{\ast}=0.09$) \cite{rice} and Zwierlein {\it et al}. ($P_c = 0.71$) \cite{mit, private} at the same resonance field ($H=832 {\rm G}$). The former (latter)  is BCS-FFLO (FFLO-normal) transition.

This phase diagram shown in Fig.4 is quite generic, which describes various physical systems, such as CDW, SDW, or stripe phase in high $T_c$ superconductors \cite{stripe}, but usually expressed in terms of $T$ vs $\delta \mu$. Here the population imbalance is a control parameter of the system. This phase diagram in $T$ vs $P$ is qualitatively same for other coupling strengths or $\Delta_0 (0)/E_{F} (0)$ or $1/k_F|a|$. Figure 5 displays a schematic phase diagram in $T$, $P$ and $1/k_F|a|$ of the attractive side only. In $T$ vs $1/k_F|a|$ plane the phase boundary shows $T_{c0} \propto e^{-\pi/2k_F|a|}$. The phase boundary in $P$ vs $1/k_F|a|$ plane is also an exponential behavior because $P_c\propto \Delta_0$.

\begin{figure}[b!]
\includegraphics[width=0.8\linewidth]{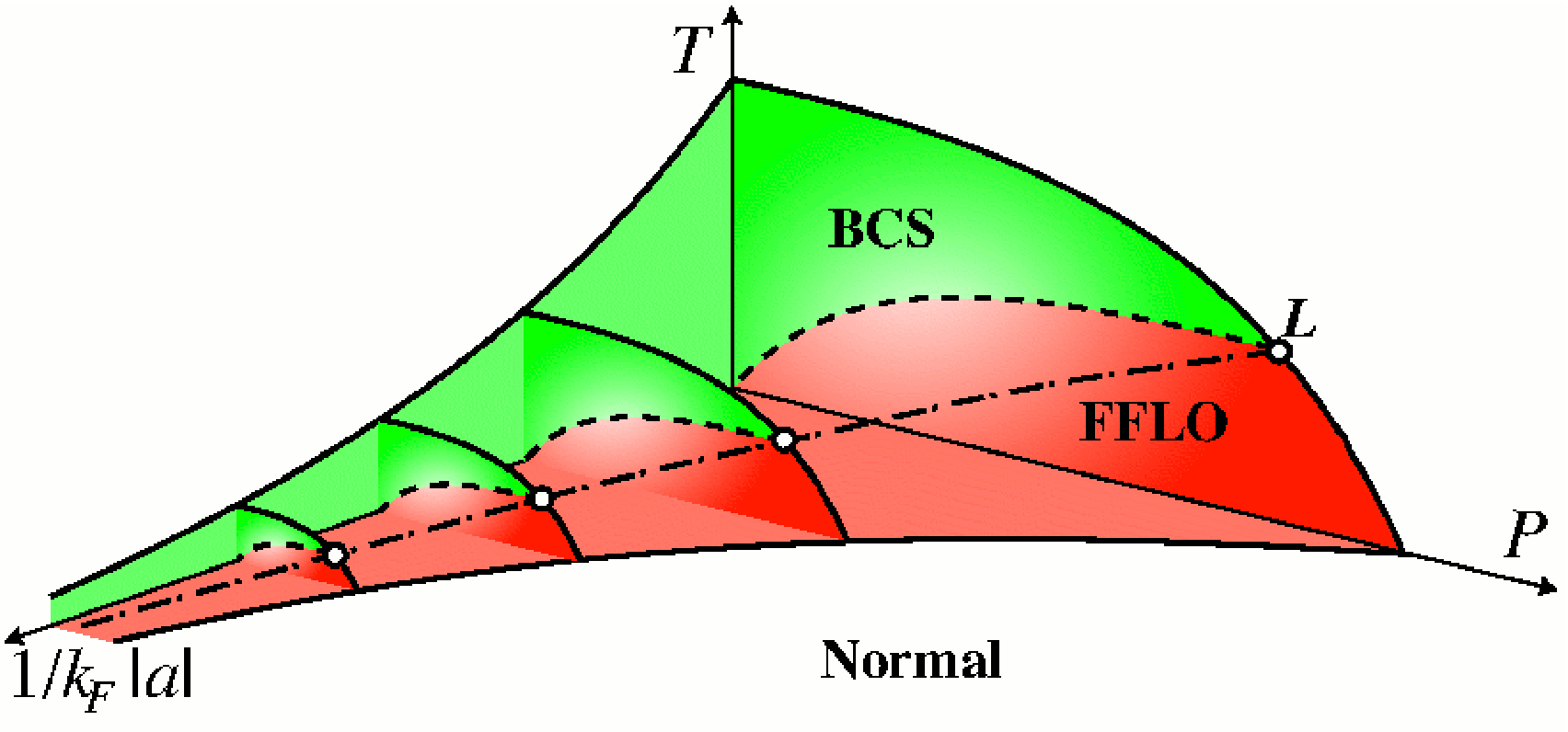} 
\caption{
(color online) Schematic phase diagram in $T$, $P$ and $1/k_F|a|$. The origin is at $T$=$P$=$1/k_F|a|$=0. 
}
\label{fig:gap}
\end{figure}

Since it is rather difficult to distinguish between the FFLO and BCS only from the density profiles, we definitely need further experimental probes to identify each phase. According to Ref.~\cite{mit} vortices in the outer region (II) and (III) become invisible while only in (I) visible vortices are sustained. The quantum depletion of particle density at a vortex core only occurs when $\Delta/E_F$ becomes large enough \cite{hayashi, vortex}. Excess majority species in region (II) fill out the vortex core preferentially, making a vortex invisible through the density profile measurement even though the phase of $\Delta({\bf r})$ winds around.

In conclusion, we have constructed a generic phase diagram in $T$ vs $P$ and explained various experimental aspects, including items ($i$)--($iii$). We have approached the strong coupling limit problem on resonance from the weak coupling using the BdG formalism by assuming that there is no phase transition between them. 
An understanding of the critical imbalance $0.7$ from first principles is an outstanding problem, which belongs to future work.

The authors thank M.W. Zwierlein for valuable discussions and R. Hulet for useful information.

\end{document}